\documentclass[twoside,11pt]{article}
\usepackage{epsfig}

\newcommand{\be}{\begin{equation}}
\newcommand{\ee}{\end{equation}}
\newcommand{\bea}{\begin{eqnarray}}
\newcommand{\eea}{\end{eqnarray}}

\usepackage{amsmath}
\usepackage{amsfonts}
\usepackage{amssymb}
\usepackage{amsxtra}

\usepackage{graphicx}
\usepackage{supertabular}
\begin{document}
\title{ \vspace{1cm} A trial to understand the supersymmetry relations 
through extension of the second  quantized fermion and boson fields,  
either to strings or to odd dimensional spaces}

\author{N.S.\ Manko\v c Bor\v stnik$^{1}$, H.B.\ Nielsen$^{2}$\\
\\
$^1$Department of Physics, University of Ljubljana\\
SI-1000 Ljubljana, Slovenia   \\
 $^2$Niels Bohr Institute, University of Copenhagen\\
 Blegdamsvej 17, Copenhagen, Denmark\\
Bled Proceedings 2024\\ 
 }

\maketitle
 

\begin{abstract}

The article studies the extension of the internal spaces of
fermion and boson second quantized fields,
described by the superposition of odd (for fermions)
and even (for bosons) products of the operators
$\gamma^ {a}$, to strings and odd dimensional spaces.\\
For any symmetry $SO(d-1,1)$ of the internal spaces, it is
the number of fermion fields (they appear in families and
have their Hermitian conjugated partners in a separate
group) equal to the number of boson fields (they appear
in two orthogonal groups), manifesting a kind of
supersymmetry, which differs from the usual supersymmetry.\\
The article searches for the supersymmetry arising from
extending the ``basis vectors'' of second  quantized
fermion and boson fields  described in $d=2(2n+1)$
(in particular $d=(13+1)$)
either to strings or to odd-dimensional spaces
($d=2(2n+1)+1$).
%
\end{abstract} 

%
\section{Introduction}
\label{introduction}
The contribution, appearing in this proceedings, with the title
``Do we understand the internal spaces of second quantized
fermion and boson fields, with gravity included?'' and in the
references therein~\cite{norma93,pikanorma2005,%
nh02,n2014matterantimatter,nd2017,2020PartIPartII,
nh2021RPPNP,n2023NPB,n2023MDPI, n2024NPB,nh2023dec},
shortly presents the properties of fermion and boson fields  
treated uniquely if they all start as massless fields.  

The {\it spin-chargge-family} theory, describing the internal 
spaces of fermion and boson second quantized fields by 
``basis vectors'' which are the superposition of products of an odd 
number of $\gamma^a$ (for 
fermions) and even  number of $\gamma^a$ (for bosons), 
requires the same number of fermion and boson 
``basis vectors''. Arranging the ``basis vectors'' to 
be the eigenvectors of the (chosen) Cartan subalgebra members 
(equal to $\frac{d}{2}$ for an even $d$)
of the Lorentz algebra in internal spaces of fermions and bosons, 
the theory offers an elegant description of the second quantized 
fermion and boson fields, explaining the second quantization 
postulates.

The author, with the collaborators~\cite{norma93,pikanorma2005,%
nh02,n2014matterantimatter,nd2017,2020PartIPartII,
nh2021RPPNP,n2023NPB,n2023MDPI, n2024NPB,nh2023dec}, 
arrange the ``basis vectors'' to be  the products of nilpotents 
($\stackrel{ab}{(k)}:=\frac{1}{2}(\gamma^a +
\frac{\eta^{aa}}{ik} \gamma^b)\,,  (\stackrel{ab}{(k)})^2=0\,$) 
and projectors ($\stackrel{ab}{[k]}:=
\frac{1}{2}(1+ \frac{i}{k} \gamma^a \gamma^b)\,, 
(\stackrel{ab}{[k]})^2=\stackrel{ab}{[k]}\,$), with the properties
$S^{ab} \,\stackrel{ab}{(k)} = \frac{k}{2} \,\stackrel{ab}{(k)}$\,,
$S^{ab}\,\stackrel{ab}{[k]} = \frac{k}{2} \,\stackrel{ab}{[k]}$\,
with $k^2=\eta^{aa} \eta^{bb}$; $S^{ab}=\frac{i}{2}\,\gamma^a
\gamma^b, a \ne b$.

``Basis vectors'' of fermions, chosen to be the algebraic products 
of an odd number of nilpotents (at least one, the rest are 
projectors),  and ``basis vectors'' of bosons, chosen to be the 
algebraic products of an even number of nilpotents (or only of 
projectors) are correspondingly eigenvectors of all  the 
$\frac{d}{2}$ $S^{ab}$ Cartan subalgebra members of one 
irreducible representation of fermions.

Fermion ``basis vectors'' appear in  $2^{\frac{d}{2}-1}$
irreducible representations --- families --- each family having 
$2^{\frac{d}{2}-1}$ members, including in $d=2(2n+1)$ 
fermions and antifermions. All the fermion ``basis vectors''  
are mutually orthogonal, while the ``basis vectors'' fulfil
together with their Hermitian conjugated partners, appearing in
a separate group, the Dirac second quantization postulates for 
fermions. Fermion ``basis vectors'' and their Hermitian 
conjugated  partners have together $2^{d-1}$ members.
 
Boson ``basis vectors'' appear in two orthogonal groups, each 
of the two groups with $2^{\frac{d}{2}-1}\times$
$2^{\frac{d}{2}-1}$ members have their Hermitian 
conjugated partners within the same group.
 There are two kinds of ``basis vectors'' of boson fields (the 
ordinary theory does not have two kinds): One kind transforms 
family members within the family (anyone), and the other 
transforms any member of a family to the same member
of another (or the same) family.

The number of fermion ``basis vectors'' is equal to the number 
of boson ``basis vectors'' (of the kind transforming family 
members within each family) manifesting a kind of 
supersymmetry, which differ from the one offered by string 
theories~\cite{Kevin,Blumenhagen}.

In this contribution, the authors  using the {\it spin-charge-family} 
theory to represent properties of fermion and boson fields, discuss 
the extension of the ``basis vectors'' in $d=2(2n+1)$ to strings and 
to odd-dimensional spaces in order to see which kind of 
supersymmetry the extension could offer.

The vacuum state is constructed from only ``basis  vectors'' of 
bosons, with spins and charges equal to zero. There are also all 
fermion and boson ``basis vectors'' present, all with the momentum 
equal to zero, if fermions and bosons are not active.

Charges and spins of the vacuum with all the ``basis vectors'' of 
fermions and bosons with no momenta present are zero. Any 
contribution to the vacuum can be written as the algebraic product 
of  a Hermitian conjugated  ``basis vector'', $({\rm``basis\,
 vector''})^{\dagger}$, algebraically, $\,*_A \,$, multiplied 
by another ``basis vector'' ($ ``basis \,vector''$). All the members 
of one family have the same contribution to the vacuum state and
each family has its own contribution to the vacuum state.\\


Knowing the ``basis vectors'' of  fermions, the theory enables to
represent all the ``basis vectors'' of both kinds of bosons as the 
algebraic products of, a ``basis vector'' $\,*_A \,({\rm a\,``basis\,
vector''})^{\dagger}$ for one kind of bosons, or as, a 
$({\rm``basis\,vector''})^{\dagger}\,*_A \,$ a ``basis \,vector'', 
for the second kind of bosons. The fermion ``basis vectors'', being 
the superposition if an odd products of $\gamma^{a}$'s, mutually 
correspondingly anti-commute, the boson ``basis vectors'', they  
are algebraic products of two  fermion ``basis vectors'', 
correspondingly commute. The Dirac's anti-commutation relations 
for fermions and commutation relations for bosons are here 
explained. 

We therefore know all the ``basis vectors'' of fermions as the 
eigenvectors of all the Cartan subalgebra members (which 
express the eigenvalues of all the spins and charges of 
fermions) as well as the ``basis vectors'' of bosons as the 
eigenvectors of all the Cartan subalgebra members (which express 
the eigenvalues of all the spins and charges of bosons).

\vspace{2mm}

To understand the properties of fermions and bosons ``basis 
vectors'', let us  start to explain the details of the proposed 
theory of N.S.M.B. by shortly repeating the Sect.~2 from 
the contribution of one of the
two authors (N.S.M.B.) in this proceedings (entitled ``Do we 
understand the internal spaces of second quantized fermion and 
boson fields, with gravity included?'').

\vspace{2mm}

%
%
\subsection{From Grassmann algebra to ``basis vectors'' describing 
the internal spaces of fermions and bosons}
\label{GrassmannClifford}
This is a short overview of Sect.~2 from the contribution of 
one of the two authors (N.S.M.B.) in this proceedings (entitled 
``Do we understand the internal spaces of second quantized 
fermion and boson fields, with gravity included?'').

Starting with the Grassmann algebra~\cite{norma93,n2023NPB}, 
offering for the description for the internal degrees of freedom of 
fermions and bosons $2\times 2^{d}$ anticommuting operators in 
$d$-dimensional space~\cite{n2019PRD},  $\theta^{a}$ and the 
derivatives with respect to $\theta^{a}$, 
$ \frac{\partial}{\partial \theta_{a}}$ ~\cite{norma93}, fulfilling 
the relations $\{\theta^{a}, \theta^{b}\}_{+}=0\,,$
$\{\frac{\partial}{\partial \theta_{a}}, 
\frac{\partial}{\partial \theta_{b}}\}_{+} =0\,,$
$\{\theta_{a},\frac{\partial}{\partial \theta_{b}}\}_{+} =\delta_{ab}\,,
\;(a,b)=(0,1,2,3,5,\cdots,d)\,$ we find two kinds of the operators
$\gamma^{a}$ and $\tilde{\gamma}^{a}$
\begin{small}
\begin{eqnarray}
\label{clifftheta1}
\gamma^{a} &=& (\theta^{a} + \frac{\partial}{\partial \theta_{a}})\,, \quad
\tilde{\gamma}^{a} =i \,(\theta^{a} - \frac{\partial}{\partial \theta_{a}})\,,\nonumber\\
\theta^{a} &=&\frac{1}{2} \,(\gamma^{a} - i \tilde{\gamma}^{a})\,, \quad
\frac{\partial}{\partial \theta_{a}}= \frac{1}{2} \,(\gamma^{a} + i \tilde{\gamma}^{a})\,,
\nonumber\\
\end{eqnarray}
offering together $2\cdot 2^d$ operators: $2^d$ are superposition of 
products of $\gamma^{a}$ and $2^d$ of $\tilde{\gamma}^{a}$,
with the properties
\begin{eqnarray}
\label{gammatildeantiher}
\{\gamma^{a}, \gamma^{b}\}_{+}&=&2 \eta^{a b}= \{\tilde{\gamma}^{a},
\tilde{\gamma}^{b}\}_{+}\,, \nonumber\\
\{\gamma^{a}, \tilde{\gamma}^{b}\}_{+}&=&0\,,\quad
(a,b)=(0,1,2,3,5,\cdots,d)\,, \nonumber\\
(\gamma^{a})^{\dagger} &=& \eta^{aa}\, \gamma^{a}\, , \quad
(\tilde{\gamma}^{a})^{\dagger} = \eta^{a a}\, \tilde{\gamma}^{a}\,.
\end{eqnarray}

 Postulating how do $\tilde{\gamma}^{a}$'s operate on $\gamma^a$'s, 
\begin{eqnarray}
\{\tilde{\gamma}^a B &=&(-)^B\, i \, B \gamma^a\}\, |\psi_{oc}>\,,
\label{tildegammareduced0}
\end{eqnarray}
with $(-)^B = -1$, if $B$ is (a function of) odd products of $\gamma^a$'s,
otherwise $(-)^B = 1$~\cite{nh02}, with the vacuum state $|\psi_{oc}>$,
the two Clifford subalgebras, $\gamma^a$ and $\tilde{\gamma}^a$ reduce
to the one described by $\gamma^a$~\cite{nh02,norma93,JMP2013},
while $\tilde{\gamma}^a$ can be used to describe the quantum numbers
of the irreducible representations of the superposition of odd products of
$\gamma^a$; $S^{ab}=\frac{i}{2} \gamma^{a}\gamma^{b}, a\ne b,$ determine quantum numbers of the members of one (anyone) irreducible representation, while $\tilde{S}^{ab}=\frac{i}{2} \tilde{\gamma}^a \tilde{\gamma}^b, a\ne b,$ determine quantum numbers of
each of the irreducible representations. Each irreducible representation is 
called a family. We recognize that
the quantum numbers of the two kinds of the superposition of even
products of $\gamma^a$ are determined by ${\cal S}^{ab} = (S^{ab}
+ \tilde{S}^{ab})$.

Since the product of an odd number of  $\gamma^a$'s anti-commute 
with any other product of an odd number of  $\gamma^a$'s if all 
$\gamma^a$'s are different, while the product of an even number of  
$\gamma^a$'s commute with any other product of an odd number of  
$\gamma^a$'s or any other product of an even number of 
$\gamma^a$'s if all $\gamma^a$'s are different, we use the 
superposition of the odd products of $\gamma^a$'s to describe the 
internal spaces of fermion fields, and the superposition of the even 
products of $\gamma^a$'s to describe
the internal spaces of boson fields. 

We call the superposition of products of $\gamma^a$'s, used to 
determine internal space of fermions and bosons ``basis vectors'':
Odd ``basis vectors'' are used to determine the internal spaces 
of fermion fields when applying the vacuum state,
$|\psi_{oc}> *_{T} |0_{\vec{p}}>$.  Even ``basis vectors'' are 
used to determine the internal spaces of boson fields. 
Both kinds of ``basis vectors'' fulfil the Dirac's postulates for 
 fermion and boson fields.

%
\end{small}

In order to determine the quantum numbers of the internal spaces of
fermion and boson second quantized fields, it is useful to arrange all 
the ``basis vectors'' to be the eigenstates of the Cartan subalgebra 
members of the Lorentz algebra in internal space of fermions and 
bosons, 
\begin{small}
\begin{eqnarray}
&&S^{03}, S^{12}, S^{56}, \cdots, S^{d-1 \;d}\,, \nonumber\\
&&\tilde{S}^{03}, \tilde{S}^{12}, \tilde{S}^{56}, \cdots,  \tilde{S}^{d-1\; d}\,, 
\nonumber\\
&&{\cal {\bf S}}^{ab} = S^{ab} +\tilde{S}^{ab}\,, 
\label{cartangrasscliff}
\end{eqnarray}
\end{small}
and write the ``basis vectors'', describing the internal spaces of fermion and 
boson second quantized fields, to be the products of nilpotents and 
projectors
\begin{small}
\begin{eqnarray}
\label{nilproj}
\stackrel{ab}{(k)}:&=&\frac{1}{2}(\gamma^a +
\frac{\eta^{aa}}{ik} \gamma^b)\,, \;\;\; (\stackrel{ab}{(k)})^2=0\, , 
\nonumber \\
\stackrel{ab}{[k]}:&=&
\frac{1}{2}(1+ \frac{i}{k} \gamma^a \gamma^b)\,, \;\;\;
(\stackrel{ab}{[k]})^2=\stackrel{ab}{[k]}.
\end{eqnarray}
\end{small}
Each nilpotent and projector is chosen to be the eigenvector of one of 
$\frac{d}{2}$ (for $d$ even) members of the Cartan subalgebra.

The products of an odd number of nilpotents anti-commute, 
correspondingly: at least one nilpotent is needed so that a ``basis vector'' describes the internal space of a fermion field, the rest are projectors. 
They appear in $2^{\frac{d}{2}-1}$ irreducible representations,  representing 
families; each family has $2^{\frac{d}{2}-1}$ members  which are 
obtainable from any other member by $S^{ab}$; the family member 
of any other family is obtainable by $\tilde{S}^{ab}$ which determine 
the quantum numbers  of a family. 

The Hermitian conjugated partners of nilpotents belong to a different 
group of $ 2^{\frac{d}{2}-1}$ members in $2^{\frac{d}{2}-1}$ families. 
We call the objects of an odd number of nilpotents, offering the ``basis
vectors'' of fermion fields, $ \hat{b}^{m \dagger}_{f } $. These ``basis
vectors'' together with their Hermitian conjugated partners, 
$ (\hat{b}^{m \dagger}_{f})^{\dagger}$ =$ \hat{b}^{m}_{f} $,
fulfil the postulates for the second quantized fermion fields, 
when applying on the vacuum state, $|\psi_{oc}>= 
\sum_{f=1}^{2^{\frac{d}{2}-1}}\,\hat{b}^{m}_{f}{}_{*_A}
\hat{b}^{m \dagger}_{f} \,|\,1\,>$, with $m$ any of the members.

All the odd ``basis vectors'' are orthogonal among themselves, and all 
the members of their Hermitian conjugated partners are orthogonal 
among themselves, 
\begin{eqnarray}
\hat{b}^{m \dagger}_f *_{A} \hat{b}^{m `\dagger }_{f `}&=& 0\,,
\quad \hat{b}^{m}_f *_{A} \hat{b}^{m `}_{f `}= 0\,, \quad \forall m,m',f,f `\,.
\label{orthogonalodd}
\end{eqnarray}

\vspace{2mm}

The products of an even number of nilpotents commute. They appear 
in two orthogonal groups, ${}^I\hat{{\cal A}}^{m \dagger}_{f}$ and 
${}^{II}\hat{{\cal A}}^{m \dagger}_{f}$ ,
each group has $ 2^{\frac{d}{2}-1} \times$ 
$2^{\frac{d}{2}-1}$ members with the Hermitian conjugated partners
within the same group. They fulfil the postulates of Dirac for the second 
quantized boson fields. Their eigenvalues of the Cartan subalgebra 
members, ${\cal S}^{ab}=$ $(\tilde{S}^{ab}+ S^{ab})$, are 
correspondingly, $\pm i$ or $\pm 1$ or $0$ (as we shall see in what 
follows). 

We have namely for nilpotents and projectors
\begin{small}
\begin{eqnarray}
\label{signature0}
S^{ab} \,\stackrel{ab}{(k)} = \frac{k}{2} \,\stackrel{ab}{(k)}\,,\quad && \quad
\tilde{S}^{ab}\,\stackrel{ab}{(k)} = \frac{k}{2} \,\stackrel{ab}{(k)}\,,\nonumber\\
S^{ab}\,\stackrel{ab}{[k]} = \frac{k}{2} \,\stackrel{ab}{[k]}\,,\quad && \quad
\tilde{S}^{ab} \,\stackrel{ab}{[k]} = - \frac{k}{2} \,\,\stackrel{ab}{[k]}\,,
\end{eqnarray}
\end{small}
with $k^2=\eta^{aa} \eta^{bb}$. We read that the eigenvalues of 
$S^{ab}$ on nilpotents are $\pm \frac{i}{2}$ or $\pm \frac{1}{2}$, and
the eigenvalues of $\tilde{S}^{ab}$ are as well $\pm \frac{i}{2}$ or 
$\pm \frac{1}{2}$, while the eigenvalues of 
$S^{ab}$ on projectors are $\pm \frac{i}{2}$ or $\pm \frac{1}{2}$ and
the eigenvalues of $\tilde{S}^{ab}$ on projectors are $\mp \frac{i}{2}$ 
or $\mp \frac{1}{2}$.
Correspondingly are the applications of ${\cal S}^{ab}=(\tilde{S}^{ab}+
S^{ab})$ on nilpotents equal to $\pm i$ or $\pm 1$, while  the 
applications of ${\cal S}^{ab}=(\tilde{S}^{ab}+S^{ab})$ on any 
projector give zero.

The Clifford even ``basis vectors'' belonging to two different groups are 
orthogonal, 
\begin{eqnarray}
\label{AIAIIorth}
{}^{I}{\hat{\cal A}}^{m \dagger}_{f} *_A {}^{II}{\hat{\cal A}}^{m \dagger}_{f}
&=&0={}^{II}{\hat{\cal A}}^{m \dagger}._{f} *_A
{}^{I}{\hat{\cal A}}^{m \dagger}_{f}\,.
\end{eqnarray}
The members of each of these two groups have the property.
\begin{eqnarray}
\label{ruleAAI}
{}^{i}{\hat{\cal A}}^{m \dagger}_{f} \,*_A\, {}^{i}{\hat{\cal A}}^{m' \dagger}_{f `}
\rightarrow \left \{ \begin{array} {r}
{}^{i}{\hat{\cal A}}^{m \dagger}_{f `}\,, i=(I,II) \\
{\rm or \,zero}\,.
\end{array} \right.
\end{eqnarray}

\vspace{2mm}
Let us repeat:
Half of $2^d$ different products of $\gamma^a$'s are odd, and half of them are 
even, manifesting a kind of ``supersymmetry'',  which distinguishes from the ordinary supersymmetry.

\vspace{2mm}
 
 The algebraic application, $*_{A}$, of even ``basis vectors''
${}^{I}{\hat{\cal A}}^{m \dagger}_{f }$ on odd ``basis vectors''
$ \hat{b}^{m' \dagger}_{f `} $ and the odd ``basis vectors''
$ \hat{b}^{m \dagger}_{f } $ on ${}^{II}{\hat{\cal A}}^{m \dagger}_{f }$
 gives%
\begin{eqnarray}
\label{calIAbbIIA1234gen}
{}^{I}{\hat{\cal A}}^{m \dagger}_{f } \,*_A \, \hat{b}^{m' \dagger }_{f `}
\rightarrow \left \{ \begin{array} {r} \hat{b }^{m \dagger}_{f `}\,, \\
{\rm or \,zero}\,,
\end{array} \right.\\
\hat{b}^{m \dagger }_{f } *_{A} {}^{II}{\hat{\cal A}}^{m' \dagger}_{f `} \,
\rightarrow \left \{ \begin{array} {r} \hat{b }^{m \dagger}_{f ``}\,, \\
{\rm or \,zero}\,,
\end{array} \right.
\end{eqnarray}
while
\begin{eqnarray}
\label{calbIA1234gen}
\hat{b}^{m \dagger }_{f } *_{A} {}^{I}{\hat{\cal A}}^{m' \dagger}_{f `} = 0
\,, \quad
{}^{II}{\hat{\cal A}}^{m \dagger}_{f } \,*_A \, \hat{b}^{m' \dagger }_{f `}= 0
\,,\;\;
\forall (m, m`, f, f `)\,.
\end{eqnarray}

If we know the odd ``basis vectors'' $\hat{b}^{m \dagger }_{f }$, we are 
able to generate all the 
Clifford even $ {}^{i}{\hat{\cal A}}^{m' \dagger}_{f `}, i=(I,II)$ 
``basis vectors'' 

\begin{eqnarray}
\label{AIbbdagger}
{}^{I}{\hat{\cal A}}^{m \dagger}_{f}&=&\hat{b}^{m' \dagger}_{f `} *_A 
(\hat{b}^{m'' \dagger}_{f `})^{\dagger}\,.
\end{eqnarray}
\begin{eqnarray}
\label{AIIbdaggerb}
 {}^{II}{\hat{\cal A}}^{m \dagger}_{f}&=&
(\hat{b}^{m' \dagger}_{f `})^{\dagger} *_A 
\hat{b}^{m' \dagger}_{f `'}\,. 
\end{eqnarray}

We overviewed so far the properties of the internal spaces of fermion and 
boson second quantized fields. Describing the second quantized fermion 
and boson fields with nonzero momenta in $d=(3+1)$, we represent 
fermion and boson fields by a tensor product, $*_{T}$, of the ``basis 
vectors'' representing internal spaces and the basis in ordinary space,
$\hat{b}^{\dagger}_{\vec{p}}\;, $~(\cite{n2024NPB}, App. D),
\begin{eqnarray}
\label{creatorp}
|\vec{p}>&=& \hat{b}^{\dagger}_{\vec{p}} \,|\,0_{p}\,>\,,\quad
<\vec{p}\,| = <\,0_{p}\,|\,\hat{b}_{\vec{p}}\,, \nonumber\\
<\vec{p}\,|\,\vec{p}'>&=&\delta(\vec{p}-\vec{p}')=
<\,0_{p}\,|\hat{b}_{\vec{p}}\; \hat{b}^{\dagger}_{\vec{p}'} |\,0_{p}\,>\,,
\nonumber\\
&&{\rm pointing \;out\;} \nonumber\\
<\,0_{p}\,| \hat{b}_{\vec{p'}}\, \hat{b}^{\dagger}_{\vec{p}}\,|\,0_{p}\, > &=&\delta(\vec{p'}-\vec{p})\,,
\end{eqnarray}
with the normalization $<\,0_{p}\, |\,0_{p}\,>=1$.\\

The fermion creation operators for a free massless fermion field 
of the energy $p^0 =|\vec{p}|$, belonging to a family $f$ and to a 
superposition of family members $m$ applying on the extended 
vacuum state including both spaces, $|\psi_{oc}>\,*_{T}\, |0_{\vec{p}}>$,  
can be expressed as
\begin{small}
\begin{eqnarray}
\label{wholespacefermions}
{\bf \hat{b}}^{s \dagger}_{f} (\vec{p}) \,&=& \,
\sum_{m} c^{sm}{}_f (\vec{p}) \,\hat{b}^{\dagger}_{\vec{p}}\,*_{T}\,
\hat{b}^{m \dagger}_{f} \, \,.
\end{eqnarray}
\end{small}
The creation operators $ \hat{\bf b}^{s\dagger}_{f }(\vec{p}) $ and their
Hermitian conjugated partners annihilation operators
$\hat{\bf b}^{s}_{f }(\vec{p}) $, creating and annihilating the single 
fermion states, respectively, fulfil when applying the vacuum state,
$|\psi_{oc}> *_{T} |0_{\vec{p}}>$, the anti-commutation relations for 
the second quantized fermions, postulated by Dirac (Ref.~\cite{n2023NPB},
Sect.3), explaining the Dirac's second quantization postulates for fermions, 
Eq.~(28) in this proceedings of the author N.S.M.B..\\

The even ``basis vectors'' have to carry the space index $\alpha$ which is
 equal to $\mu=(0,1,2,3)$ if they describe the vector component of the ``basis
vectors'', and they are equal to $\sigma=(5,6,...)$ if describing the scalar
components of the ``basis vectors''
\begin{eqnarray}
\label{wholespacebosons}
{\bf {}^{i}{\hat{\cal A}}^{m \dagger}_{f \alpha}} (\vec{p}) \,&=&
{}^{i}{\hat{\cal C}}^{ m}{}_{f \alpha} (\vec{p})\,*_{T}\,
{}^{i}{\hat{\cal A}}^{m \dagger}_{f} \, \,, i=(I,II)\,,
\end{eqnarray}
with ${}^{i}{\hat{\cal C}}^{ m}{}_{f \alpha} (\vec{p})=
{}^{i}{\cal C}^{ m}{}_{f \alpha}\,\hat{b}^{\dagger}_{\vec{p}}$, with
$\hat{b}^{\dagger}_{\vec{p}}$ defined in Eqs.~(\ref{creatorp}). 
We treat free massless bosons of momentum $\vec{p}$ and energy 
$p^0=|\vec{p}|$ and of particular ``basis vectors'' 
${}^{i}{\hat{\cal A}}^{m \dagger}_{f}$'s which are the eigenvectors 
of all the Cartan subalgebra members.
The creation operators for boson gauge fields commute, explaining the 
Dirac's second quantization postulates for bosons.\\

In Table 10.4 in the contribution of the author N.S.M.B. in this 
proceedings, (or Table 4. in ~\cite{n2023NPB}) we can find how do
one family of the ``basis vectors'' of quarks and leptons and 
anti-quarks and anti-leptons look like.  
The {\it spin-charge-family} requires the existence of the right-handed 
neutrinos and left-handed anti-neutrinos.  We can read  that the
$SO(7,1)$ content of $SO(13,1)$ are the same for quarks and leptons, 
and the same for anti-quarks and anti-leptons. Quarks and leptons 
differ only in the last product, in the $SU(3)\times U(1)$ content.

%

The quantum numbers of $ u_{L}^{c1}$, presented in Table  10.4 
(in the contribution of the author N.S.M.B. in this proceedings) in the 
seventh line 
$ u_{L}^{c1}$$(\equiv \stackrel{03}{[-i]}\,\stackrel{12}{[+]}|
\stackrel{56}{[+]}\,\stackrel{78}{[-]}||\stackrel{9 \;10}{(+)}
\;\;\stackrel{11\;12}{[-]}\;\;\stackrel{13\;14}{[-]}$, are:
$S^{12}$ $=\frac{1}{2}$, $S^{03}$ $=-\frac{i}{2}$,
$\tau^{13}=\frac{1}{2}(S^{56}-S^{78})=
\frac{1}{2}$, $\tau^{23}=\frac{1}{2}(S^{56}+S^{78})=0$,
$\tau^{33}=\frac{1}{2}(S^{9\, 10}- S^{11\,12})=
\frac{1}{2}$, $\tau^{38}=\frac{1}{2\sqrt{3}}(S^{9\, 10}+
S^{11\,12}- 2 S^{13\,14})=\frac{1}{2\sqrt{3}}$, the {\it fermion}
quantum number $\tau^4=- \frac{1}{3}(S^{9\, 10}+
S^{11\,12}+ S^{13\,14})= \frac{1}{6}$, $Y = (\tau^{23} + 
\tau^4)=\frac{1}{6}$, $Q=(\tau^{13} +Y)=\frac{2}{3}$.\\
The photon ``basic vector''
${}^{II}{\hat{\cal A}}^{ \dagger}_{ph u_{L}^{c1 }
 \rightarrow u_{L}^{c1 }}$ can be found by the
multiplication of $ u_{L}^{c1}$ from the left-hand side
by $\gamma^9$: $\gamma^9$ $ u_{L}^{c1}\rightarrow$
$\stackrel{03}{[-i]}\,\stackrel{12}{[+]}|\stackrel{56}{[+]}
\,\stackrel{78}{[-]}||\stackrel{9 \;10}{[-]}
\;\;\stackrel{11\;12}{[-]}\;\;\stackrel{13\;14}{[-]}$.
 The photon ``basis vector'' ${}^{II}{\hat{\cal A}}^{ \dagger}_{ph u_{L}^{c1 }
 \rightarrow u_{L}^{c1 }}
(\equiv \stackrel{03}{[-i]}\,\stackrel{12}{[+]}|\stackrel{56}{[+]}
\,\stackrel{78}{[-]}||\stackrel{9 \;10}{[-]}
\;\;\stackrel{11\;12}{[-]}\;\;\stackrel{13\;14}{[-]}$), having
all the members of the algebraic
product equal to projectors, which obey for even ``basis 
vectors'' the relation ${\cal S}^{ab}= (S^{ab} +
\tilde{S}^{ab})$, and has correspondingly all the
quantum numbers equal zero, can not change the 
internal space quantum numbers of an odd ``basis 
vectors'', what photons do not. Photons can give to 
fermions only the momentum in the ordinary 
space-time.\\

Let us point out again that knowing all the odd ``basis vectors''
describing the internal spaces of fermions we are able to write 
all the even ``basis vectors'' describing two groups of bosons,
Eqs.~(\ref{AIbbdagger}, \ref{AIIbdaggerb}). 

The photon ``basis vector'' ${}^{II}{\hat{\cal A}}^{ \dagger}_{ph u_{L}^{c1 }
 \rightarrow u_{L}^{c1 }}
(\equiv \stackrel{03}{[-i]}\,\stackrel{12}{[+]}|\stackrel{56}{[+]}
\,\stackrel{78}{[-]}||\stackrel{9 \;10}{[-]}
\;\;\stackrel{11\;12}{[-]}\;\;\stackrel{13\;14}{[-]}$) can be written 
as $(u_{L \, 7^{th}}^{c1 })^{\dagger}\,*_{A}\, u_{L \, 7^{th}}^{c1 }$ or as
$(\bar{u}^{\bar{c1}}_{R\, 39^{th}})^{\dagger}\,*_{A}\, 
\bar{u}^{\bar{c1}}_{R\, 39^{th}}$ .

The photon  
 ${}^{I}{\hat{\cal A}}^{ \dagger}_{ph \,\bar{u}^{\bar{c1}}_{R}
 \rightarrow \bar{u}^{\bar{c1}}_{R}}$ can be represented as 
 $\bar{u}^{\bar{c1}}_{R\, 39^{th}}\,*_A\,
 (\bar{u}^{\bar{c1}}_{R\, 39^{th}})^{\dagger}$.
 
 One can find in Eqs.~(22,23,24) of the contribution of the author 
 N.S.M.B. in this proceedings the ``basis vectors''
 for gravitons (${}^{I}{\hat{\cal A}}^{ \dagger}_{gr\, u^{c1}_{R\uparrow}
\rightarrow u^{c1}_{R\downarrow}} = u^{c1}_{R\, 2^{nd}}
\,*_A\, (u^{c1}_{R\, 1^{st}})^{\dagger}$), weak bosons
(${}^{I}{\hat{\cal A}}^{ \dagger}_{w1 \,d^{c1} _{L}
\rightarrow u^{c1}_{L}}= u^{c1}_{L\, 7^{th}}\, *_A\,
(d^{c1}_{L\, 5 {th}})^{\dagger}$) and gluons 
(${}^{I}{\hat{\cal A}}^{ \dagger}_{gl\, d^{c1}_{L}
\rightarrow d^{c3}_{L}}= d^{c3}_{L\,21^{st}}\, *_A\,
( d^{c1}_{L\,5^{th}})^{\dagger}$).

In all these cases the numbers, $1^{st}, 5^{th}, 7^{th}, 21^{st}$ and 
$39^{th}$ tell the lines in Table 10.4 of the contribution of the author 
 N.S.M.B. in this proceedings where the odd 
``basis vectors'' of quarks, leptons, antiquarks and anti-leptons are
presented.

\section{Extensions of points in ordinary space time to strings, 
extensions of ``basis vectors'' to odd-dimensional spaces.}
\label{renormalisability}

 The description of the internal spaces of fermion and boson second 
 quantized fields with the ``basis vectors'' which are products of 
 an odd and  an even number of nilpotents, the rest are projectors,
 all are eigenvectors of the Cartan subalgebra members,
offers an equal number of fermion and  boson ``basis vectors'',
demonstrating a (kind of) supersymmetry.
Both, the extension of points in ordinary space-time to strings, and
the extension of ``basis vectors'' to odd-dimensional spaces, might 
help (following the literature~\cite{Kevin,Blumenhagen,FadeevPopov,%
n2023MDPI}) to achieve renormalizability of the proposed 
{\it spin-charge-family} theory.

To extend the points in ordinary space-time to strings, we must define
the ``basis vectors'' on a string with coordinates $(\sigma, \tau)$.

We have, in this case, two odd and two even ``basis vectors'' which are
the eigenvectors of the Cartan subalgebra members with the eigenvalues
for nilpotents $S^{01}=\pm \frac{i}{2}, \tilde{S}^{01}=
\pm \frac{i}{2}$, and for the projectors ${\cal S}^{01}=
(S^{01}+\tilde{S}^{01})=0$, Eq.~(\ref{nilproj}).
\begin{small} 
 \begin{eqnarray}
 \label{1+1oddeven}
 && {\rm  \;Clifford \;odd}\nonumber\\
 \hat{b}^{ 1 \dagger}_{1 s}&=&\stackrel{01}{(+i)_s}\,, \quad 
 \hat{b}^{ 1 }_{1s}=\stackrel{01}{(-i)_s}\,,\nonumber\\
 &&{\rm \;Clifford \;even} \;\nonumber\\
 {}^{I}{\bf {\cal A}}^{1 \dagger}_{1s}&=&\stackrel{01}{[+i]_s}\,, \quad 
{}^{II}{\bf {\cal A}}^{1 \dagger}_{1s}=\stackrel{01}{[-i]_s}\,.
 \end{eqnarray}
 \end{small}
 Index ${}_s$ points out that the string is concerned.
The two nilpotent``basis vectors'' are Hermitian conjugated
to each other. Making a choice that  $\hat{b}^{ 1 \dagger}_{1}=
\stackrel{01}{(+i)_s}$ is the ``basis vector'',  the second 
odd object is then its Hermitian conjugated partner. \\
The vacuum state is for this choice equal to
$|\psi_{oc_s}>=\stackrel{01}{[-i]_s}|\,1>=
(\stackrel{01}{(+i)_s})^{\dagger} \,*_A\,\stackrel{01}{(+i)_s}|\,1>$.
There is only one family
($2^{\frac{d}{2}-1}=1$) with one member ($2^{\frac{d}{2}-1}=1$) 
and one Hermitian conjugated partner.
The eigenvalue $S^{01}$ of $\hat{b}^{ 1 \dagger}_{1 s} 
\equiv \stackrel{01}{(+i)_s})$ is $\frac{i}{2}$.

Each of the two Clifford even ``basis vectors'' is self adjoint
($({}^{I,II}{\bf {\cal A}}^{1 \dagger}_{1s})^{\dagger}=$
${}^{I,II}{\bf {\cal A}}^{1 \dagger}_{1s}$), with the eigenvalues
${\cal S}^{01}=(S^{01}+\tilde{S}^{01})$ equal  to $0$ (since 
$S^{01}\stackrel{01}{[\pm i]_s}=\pm i \stackrel{01}{[\pm i]_s}$ and
$\tilde{S}^{01}\stackrel{01}{[\pm i]_s}=\mp i \stackrel{01}{[\pm i]_s}$).
 It follows that  
$$ {}^{I}{\bf {\cal A}}^{1 \dagger}_{1s}=  \hat{b}^{ 1 \dagger}_{1s}\,*_A\,
(\hat{b}^{ 1 \dagger}_{1s})^{\dagger}\, ,\quad\,\quad
{}^{II}{\bf {\cal A}}^{1 \dagger}_{1s} = (\hat{b}^{ 1 \dagger}_{1s})^{\dagger}\,*_A\, \hat{b}^{ 1 \dagger}_{1s}.$$

To find the ``basis vectors'' for second quantized fermion and 
boson fields extended to strings, we need to make a tensor 
product ,$*_{T \,`}$, of ``basis vectors'' of internal space in $d=2(2n+1)$ and 
``basis vectors'' on a string.

The extension to strings will be discussed in Subsect.~\ref{string}.

\vspace{2mm}

We can achieve a kind of a supersymmetric 
partners to the ``basis vectors'' presented fermions and 
bosons in $2(2n+1)$-dimensional internal spaces of
fermions and bosons in an odd dimensional space $d=2(2n+1) +1$.
We can find in this case two groups of ``basis vectors''~\cite{n2023MDPI}:
One group determines the anti-commuting ``basis vectors'' of 
$2^{\frac{d-1}{2}-1}$ fermions appearing in 
$2^{\frac{d-1}{2}-1}$ families, with their $2^{\frac{d-1}{2}-1}$
$\times2^{\frac{d-1}{2}-1}$ Hermitian conjugated partners 
appearing in a separate group, as well as two 
orthogonal groups each with $2^{\frac{d-1}{2}-1}$
$\times2^{\frac{d-1}{2}-1}$ of ``basis vectors'',  with their
Hermitian conjugated partners within the same group.

The second group determines anti-commuting ``basis 
vectors''  appearing in two separate orthogonal groups each 
with $2^{\frac{d-1}{2}-1}$ $\times2^{\frac{d-1}{2}-1}$ of 
``basis vectors'',  with their Hermitian conjugated 
partners within the same group, as well as the commuting 
``basis vectors'' of ``fermions'' appearing in families with 
their Hermitian conjugated partners in a separate group.

This kind of a supersymmetry will be discussed in Subsect.~\ref{2n+1}.

Both kinds of searching for the renormalizability need 
further discussions, on which we are not yet really prepared. 
More work has to be done.


%
\subsection{ Extension of ``basis vectors'' in $d=2(2n+1)$ to strings}
\label{string}

We might define the ``basis vector'' of a gravitino as a tensor product,
$\,*_{T \, `}\,$, of a photon 
 ``basis vector'' ${}^{I}{\hat{\cal A}}^{ \dagger}_{ph u_{L}^{c1 }
 \rightarrow u_{L}^{c1 }}
(\equiv \stackrel{03}{[-i]}\,\stackrel{12}{[+]}|\stackrel{56}{[+]}
\,\stackrel{78}{[-]}||\stackrel{9 \;10}{[+]}
\;\;\stackrel{11\;12}{[-]}\;\;\stackrel{13\;14}{[-]})$ (which has spins 
and charges in internal space equal to zero), for example, with 
$\hat{b}^{ 1 \dagger}_{1 s} (\equiv \stackrel{01}{(+i)_s})$ on a 
string: $\hat{b}^{ 1 \dagger}_{1 gravitino} (\equiv \stackrel{03}{[-i]}\,
\stackrel{12}{[+]}|\stackrel{56}{[+]} \,
\stackrel{78}{[-]}||\stackrel{9 \;10}{[+]}\;\;\stackrel{11\;12}{[-]}
\;\;\stackrel{13\;14}{[-]}\,*_{T\, `}\,\stackrel{01}{(+i)_s})$.
This is an anti-commuting object and could manifest gravitino if 
the photon ``basis vector'' 
${}^{I}{\hat{\cal A}}^{ \dagger}_{ph u_{L}^{c1 }
 \rightarrow u_{L}^{c1 }}$ is in a tensor product with basis in
ordinary space-time, carrying the space index $\mu=(0,1,2,3)$.

The extensions of all the other ``basis vectors''  --- either the 
ones with an odd number of nilpotents describing the internal 
spaces of fermions, or with an even number of nilpotents 
describing the internal spaces of bosons ---  by the tensor 
product, $\,*_{T\, '}\,$, with the two commuting self adjoint 
``basis vectors'' describing the internal space on the string, 
$ {}^{i}{\bf {\cal A}}^{1 \dagger}_{1 s}, i=(I,II)$, do not 
change commutation properties: The extended ``basis vectors'' 
keep commutation properties  of the ``basis vectors'' of fermions 
and bosons.

The extensions of  the ``basis vectors''  describing fermions and 
bosons by the tensor product, $\,*_{T\, '}\,$, with the nilpotent
$\hat{b}^{ 1 \dagger}_{1 s}$ do change the commutation
relations: The commuting ones become anti-commuting, the 
anti-commuting become commuting.

Let us try to see general properties of tensor products, $\,*_{T\, '}\,$,
of the ``basis vectors'' with an odd number of nilpotents 
(describing the internal spaces of the second quantized fermion 
fields) $\hat{b}^{m \dagger}_{f}$ and of the ``basis vectors'' 
with an even number of nilpotents (describing the internal 
spaces of the second quantized boson fields) 
$ {}^{I,II}{\bf {\cal A}}^{m \dagger}_{f}$ with the
``basis vectors'' of a string.\\
\vspace{2mm}

There are four possibilities:\\

{\bf i.\;\;\;}
$$ \hat{b}^{m \dagger}_{f}\, \,*_{T\, '}\,
{}^{I}{\bf {\cal A}}^{1 \dagger}_{1 s}$$ 
represents the anti-commuting ``basis vectors'' extended with 
a string offering the description of the internal spaces of fermions 
in $d=2(2n+1)$.\\

{\bf ii.\;\;\;}
$$ {}^{I,II}{\bf {\cal A}}^{m \dagger}_{f}\,*_{T \,'}\,
{}^{I}{\bf {\cal A}}^{1 \dagger}_{1 s}$$ 
represents the commuting ``basis vectors'' extended with 
a string 
offering the description of the internal spaces of bosons in 
$d=2(2n+1)$.
Since  ${}^{II}{\bf {\cal A}}^{1 \dagger}_{1 s}$ defines the 
vacuum state $|\psi_{oc s}>=\stackrel{01}{[-i]_s} |\,  1>$ for
$\hat{b}^{1 \dagger}_{1 s}$, only 
${}^{I}{\bf {\cal A}}^{1 \dagger}_{1 s}$ is used in a tensor 
product  $\,*_{T \,'}\,$.\\
\vspace{2mm}

{\bf iii.\;\;\;}
$${}^{I,II}{\bf {\cal A}}^{m \dagger}_{f}\, \,*_{T  \,'}\,
 \hat{b}^{1 \dagger}_{1 s}$$ 
represents the anti-commuting ``basis vectors'' extended with 
a string offering the description of the internal spaces of  
anti-commuting objects with the quantum numbers of bosons 
in $d=2(2n+1)$.\\
\vspace{2mm}

{\bf iv.\;\;\;}
$$ \hat{b}^{m \dagger}_{f}\, \,*_{T\, '}\, 
\hat{b}^{1 \dagger}_{1 s}$$ 
represents the commuting ``basis vectors'' extended with 
a string offering the description of the internal spaces of  
bosons with the quantum numbers of fermions in $d=2(2n+1)$.\\

\vspace{2mm}

We recognize the supersymmetry: \\


Each $ {}^{I,II}{\bf {\cal A}}^{m \dagger}_{f}$
\, $\,*_{T '}\,$  ${}^{I}{\bf {\cal A}}^{1 \dagger}_{1 s}$ and 
each $\hat{b}^{m \dagger}_{f}\, \,*_{T '}\,
{}^{I}{\bf {\cal A}}^{1 \dagger}_{1 s}$
 has a  supersymmetric partner in either 
 $ {}^{I,II}{\bf {\cal A}}^{m \dagger}_{f}\,
\,*_{T '}\, \hat{b}^{1 \dagger}_{1 s}$ or in
$ \hat{b}^{m \dagger}_{f}$ 
$\,*_{T '}\,$  $\hat{b}^{1 \dagger}_{1 s}$. 

The extension of the $2^{\frac{d}{2}-1}$ ``basis vectors''
with an odd number of nilpotents appearing in $2^{\frac{d}{2}-1}$ 
families with their Hermitian conjugated partners in a separate 
group, and of the ``basis vectors'' of an even number of nilpotents
appearing in two orthogonal groups, in a tensor extension by 
$\hat{b}^{ 1 \dagger}_{1 s}$ needs further studies to be 
understood.

\subsection{``Supersymmetry'' in odd dimensional spaces}
\label{2n+1}

Let us come to the second possibility, to find out what kind of
symmetry  the internal odd-dimensional spaces 
$d=(2(2n+1) +1)$ offer. They namely manifest two groups 
of anti-commuting ``basis vectors'' and  two groups of 
commuting ``basis vectors'', as discussed 
in the article~\cite{n2023MDPI}.
\\

``Basis vectors'' of the first part of each of the two groups have 
properties as we presented for $2(2n+1)$-dimensional spaces \\
 --- the anti-commuting ``basis vectors'' with an odd number of 
 nilpotents $ \hat{b}^{m \dagger}_{f}$ appear in families, 
their Hermitian conjugated partners form a separate group 
$\hat{b}^{m}_{f}$\\  
--- the commuting ``basis vectors'' with an even number of 
nilpotents appear in two orthogonal groups, 
$ {}^{I,II}{\bf {\cal A}}^{m \dagger}_{f}$, each group have 
the Hermitian conjugated partners  within the same group. \\

``Basis vectors'' of the second part of each of the two groups 
have completely different properties than the first part \\
--- the anti-commuting``basis vectors'' appear in 
two orthogonal groups,  with the Hermitian conjugated 
partners within the same group\\ 
--- the commuting ``basis vectors'' appear in families and 
have their Hermitian conjugated partners in a separate 
group.\\

Let us try to understand the properties of the second part 
of the ``basis vectors''. \\

These ``basis vectors'' and their Hermitian conjugated 
partners can be obtained from the first part by the application 
of $ S^{0 d}$ on the two groups of the first part.\\

 Applying  $ S^{0 d}=\frac{i}{2} \gamma^{0}\, \gamma^{d}$
(having the even number of  $\gamma^{a}$) on 
$ \hat{b}^{m \dagger}_{f}$ does not change the 
 oddness of the new object $\gamma^{0} \gamma^{d} \, 
 \hat{b}^{m \dagger}_{f}$. However, 
$\gamma^{0} \hat{b}^{m\dagger}_{f}$ represent now  
${}^{II}{\bf {\cal A}}^{m' \dagger}_{f `}$, while 
 $\gamma^{d}$  multiplying 
$ {}^{II}{\bf {\cal A}}^{m' \dagger}_{f `}$, keep the 
 oddness unchanged. \\

The application of the even operator $ S^{0 d}=\frac{i}{2} 
\gamma^{0} \gamma^{d}$ on an object with an even number 
of nilpotents ${}^{II}{\bf {\cal A}}^{m \dagger}_{f}$  does 
not change the evenness of the object 
$\gamma^{0} \gamma^{d} \, 
{}^{II}{\bf {\cal A}}^{m \dagger}_{f}$.  However, 
$\gamma^{0}\; {}^{II}{\bf {\cal A}}^{m \dagger}_{f}$ 
represent indeed $ \hat{b}^{m' \dagger}_{f `}$ while  
 $\gamma^{d}$ multiplying  $ \hat{b}^{m' \dagger}_{f `}$, 
 keep  the evenness unchanged. \\

\vspace{2mm}

We can conclude that odd dimensional spaces, $d=2(2n+1) +1$, 
 \\

{\bf i.\;\;\;} offer the anti-commuting $2^{\frac{d-1}{2}-1}$ ``basis 
vectors''  $\hat{b}^{m\dagger}_{f}$ appearing in 
$2^{\frac{d-1}{2}-1}$ families, with their Hermitian conjugated 
$2^{\frac{d-1}{2}-1}$ $\times$ $ 2^{\frac{d-1}{2}-1}$ 
partners,  $\hat{b}^{m}_{f}$, in a separate group, and\\

{\bf ii.\;\;\;} the commuting $2\times 2^{\frac{d-1}{2}-1}$
$\times$ $2^{\frac{d-1}{2}-1}$ ``basis vectors'' 
${}^{i}{\bf {\cal A}}^{m \dagger}_{f}, i=(I,II),$ appearing in two 
orthogonal groups with their Hermitian conjugated partners  
within the same group.\\

{\bf iii.\;\;\;}
They offer as well  
the  anti-commuting $2^{\frac{d-1}{2}-1}$$\times$ 
$2^{\frac{d-1}{2}-1}$ ``basis vectors'' 
${}^{i}{\bf {\cal A}}^{m \dagger}_{f}$ appearing in two 
orthogonal groups with their Hermitian conjugated
 partners within the same group, and\\
%

{\bf iv.\;\;\;} the commuting $2^{\frac{d-1}{2}-1}$ 
``basis vectors''  $\hat{b}^{m\dagger}_{f}$ appearing in 
$2^{\frac{d-1}{2}-1}$ families, with their Hermitian conjugated 
$2^{\frac{d-1}{2}-1}$ $\times$ $ 2^{\frac{d-1}{2}-1}$ 
partners,  $\hat{b}^{m}_{f}$, in a separate group.\\

Also this case needs further studies to be understood what it does 
offer.

\section{Conclusions}
\label{conclusions}

 The description  of the internal spaces of fermions and bosons
in $d=2(2n+1)$ with the ``basis vectors'' with  odd  and even 
numbers of nilpotents, respectively, offers a kind of  supersymmetry, 
existing in equal number of anti-commuting fermions and of 
commuting bosons. This is not the usual supersymmetry.

One way to achieve renormalizability of the proposed 
{\it spin-charge-family} theory might be,  following the literature~\cite{Kevin,Blumenhagen,FadeevPopov,n2023MDPI}, 
to extend the ``basis vectors'' in
$d=2(2n+1)$ with the tensor product $ \,*_{T\, '}\,$ 
with the ``basis vectors'' and Hermitian conjugated 
partners of strings, $\hat{b}^{1 \dagger}_{1 s}$ and 
 $ {}^{I}{\bf {\cal A}}^{1 \dagger}_{1 s}$:
$$ \hat{b}^{m \dagger}_{f}\, \,*_{T\, '}\,
{}^{I}{\bf {\cal A}}^{1 \dagger}_{1 s}\,, \;\;\;\;
{}^{I,II}{\bf {\cal A}}^{m \dagger}_{f}\, \,*_{T\, '}\,
{}^{I}{\bf {\cal A}}^{1 \dagger}_{1 s}\,, $$
$$  \hat{b}^{m \dagger}_{f}\, \,*_{T\, '}\,
\hat{b}^{1 \dagger}_{1 s}\,, \;\;\;\;
{}^{I,II}{\bf {\cal A}}^{m \dagger}_{f}\, \,*_{T\, '}\,
 \hat{b}^{1 \dagger}_{1 s}\,.$$ 
The second way is to extend the ``basis vectors'' in
$d=2(2n+1)$ into ``basis vectors'' in $d=(2(2n+1) +1)$. 
Again we have four possibilities:\\
 $$ {\rm The\, anti-commuting} \,\,\,\hat{b}^{m \dagger}_{f}\,, \;\;\;\; 
{\rm the \, commuting} \,\,\,{}^{I,II}{\bf {\cal A}}^{m \dagger}_{f},$$ 
$$ {\rm the\, commuting} \,\,\,\gamma^d \,\,\,\hat{b}^{m \dagger}_{f}\,, \;\;\;\; 
{\rm the\, anti-commuting} \,\,\,\gamma^{d} \,\,\,
{}^{I,II}{\bf {\cal A}}^{m \dagger}_{f}.$$
 We can conclude that each of the two possibilities, offering a kind of 
  supersymmetry, seems meaningful.
 The extension of the ``basis vectors'' in $d=2(2n+1)$ with the tensor 
 product $ \,*_{T\, '}\,$ to  strings suggests that at low enough energies only 
$ \hat{b}^{m \dagger}_{f}$\, $\,*_{T \,'}\,
{}^{I}{\bf {\cal A}}^{1 \dagger}_{1 s}$ and 
${}^{I,II}{\bf {\cal A}}^{m \dagger}_{f}$\,
 $\,*_{T \,'}\,{}^{I}{\bf {\cal A}}^{1 \dagger}_{1 s}$ can be observable. 

The extension of the ``basis vectors'' in $d=2(2n+1)$ to the
 odd-dimensional space, $d=2(2n+1) +1$,  suggests that at low enough 
 energies only  anti-commuting $\hat{b}^{m \dagger}_{f}$ and
 commuting ${}^{I,II}{\bf {\cal A}}^{m \dagger}_{f}$ can be observed. 

 Not all of them, as we realize from the observations.\\

Both suggestions need to be analysed to recognize whether any of them
leads to  a  renormalizable and anomaly-free theory.
 
It might be that nature does not need the supersymmetry ``to make the
theory renormalizable and anomaly-free.

\section*{Acknowledgments} 

The authors    thank Department of Physics, FMF, University of Ljubljana, Society of Mathematicians, Physicists and Astronomers of Slovenia,  for supporting the research on the {\it spin-charge-family} theory by offering the room and computer facilities and Matja\v z Breskvar of Beyond Semiconductor for donations, in particular for the annual workshops entitled "What comes beyond the standard models".



\begin{thebibliography}{99}
\bibitem{n2024NPB}  N. S. Manko\v c Bor\v stnik, "Can the ``basis vectors'', 
describing t he internal spaces of fermion and boson fields with the Clifford odd 
(for fermion) and Clifford even (for boson) objects, explain interactions 
among fields, with gravitons included?"  [arxiv: 2407.09482]. 
\end{thebibliography}
\end{document}